\documentstyle[]{article}
\begin{document}
\overfullrule 0 mm
\language 0
\vskip 0.5 cm
\centerline { \bf{SOME ANALYTICAL RESULTS }} 
\centerline { \bf{ON CLASSICAL TUNNELING }} 
\centerline { \bf {OF SOMMERFELD PARTICLE}} 
\vskip 0.5 cm
\centerline {\bf{ Alexander A.  Vlasov}}
\vskip 0.3 cm
\centerline {{  High Energy and Quantum Theory}}
\centerline {{Department of Physics}}
\centerline {{ Moscow State University}}
\centerline {{  Moscow, 119899}}
\centerline {{ Russia}}
\vskip 0.3cm
{\it  
A simple example of especially constructed potential barrier enables to show 
analytically (not numerically) the existence of tunneling effect for a 
Sommerfeld particle. }

03.50.De
\vskip 0.3 cm

 Let us remind that
Sommerfeld model of charged rigid sphere [1] (Sommerfeld particle) is the 
simplest theoretical model to take into consideration the "back-reaction" of 
self-electromagnetic field on  equation of motion of a radiating extended 
charged body (for point-like charge we have the known Lorentz-Dirac equation 
with all its problems: renormalization of mass, preacceleration, runaway 
solutions, etc.).

For the case of simplicity here we consider the nonrelativistic,
linear in velocity, version of Sommerfeld model [2].

So let the total charge of a  uniformly  charged sphere be $Q$,
mechanical mass - $m$, radius - $a$.  Then its equation of motion
reads [1,2]:  $$m\dot{\vec v} =\vec F_{ext}+ \eta\left[\vec v(t-2a/c) -
\vec v(t)\right] \eqno(1)$$ here  $\eta=
{Q^2 \over 3  c a^2},\ \ \vec v= d \vec R /dt,\ \ \vec R$ -
coordinate of the center of the shell, $\vec F_{ext}$ - some external force.

This equation has no runaway solutions and solutions with preacceletation.
But, as was shown in [3], it has the solutions, which one can interpret as 
 classical tunneling. The physics of such effect is simple - Sommerfeld 
particle due to retardation begins to "feel" the action of potential barrier 
too late, when this barrier is overcome.

In [3] the existence of solutions with tunneling effect was demonstrated
with the help of numerical calculations.

Here we  present one simple problem, which enables us to see the 
appearance of classical tunneling without numerical calculations but 
using only analytical considerations.

Let the Sommerfeld particle move along $z$ -axis ( $\vec R=(0,\ 0,\ R)$ ) in 
 a static electric field $\vec E =(0,\ 0, E_z)$ produced by electric 
 potential $\phi$ in a form of a step:  $$ \phi =B[\theta(z) - \theta(z-S)],\ 
 \ \ \ E_z =-{d \phi \over d z}=-B \delta(z)+B \delta(z-S)\eqno(2)$$

Then the external force $F_{ext}$, acting on rigid sphere with density of 
charge $\rho$,
$$ \rho ={Q \over 4\pi a^2}\delta(|\vec r -\vec R|-a)$$
 reads $$ F_{ext} =\int d {{\vec r}} \rho   E_z ={QB \over 4\pi 
a^2}\int d\vec r \delta(|\vec r -\vec R|-a) [-\delta(z)+\delta(z-S)]$$ With 
new variables $\vec{ r} \equiv \vec {\xi}+\vec {R},\ \ \ d\vec{ r } =d\vec{ 
\xi}= \xi^2 d \xi \sin{\theta} d\theta d\phi$, substituting 
$\cos{\theta}\equiv \mu$ (it leads to $z=R+\xi \mu$ ), integrating over 
$\phi$ and $\xi$, we get $$ F_{ext}={QB \over 2 
a}\int\limits_{-1}^{+1}d\mu [-\delta(R+a\mu)+\delta(R+a\mu-S)]= {QB \over 2 
a}\left[-\int\limits_{R-a}^{R+a} d w \delta(w) +\int\limits_{R-a-S}^{R+a-S} 
d w \delta(w)\right]$$ 
If $S\geq a$ then the expression in square brackets yields the result

$0\ \ for \ R \leq -a,\ \ -1\ for \ -a\leq R\leq +a,\ \ 0 \ 
for \ +a\leq R\leq S-a, \ \ +1\ for \ S-a\leq R\leq S+a,\ \  
0 \ for \ R\geq S+a$

After this, eq.(1) in dimensionless variables 
 $y= R/L,\ \ \tau =ct/L,\ \ \delta=2a/L$,  for the 
relations $S=L=2a$ (taken for simplicity), is reduced to 
$${d^2 y 
\over d\tau ^2} = k \left[ {d y(\tau -1) \over d\tau } -{d y(\tau ) \over 
d\tau }\right] + \lambda \Phi \eqno(3)$$ here $k={2Q^2 \over 3 m c^2 a},\ 
\ \lambda={ Q B \over m c^2}$, 

and
$$\Phi=\left\{\matrix{ 0,& y<-1/2, \cr -1,& -1/2<y<1/2, \cr +1,& 
1/2<y<3/2,\cr 0,& 3/2<y, \cr }\right.$$

Let us note that in the limit of zero particle's size 
$a\to 0$ the force $F_{ext}$ tends to two delta-functions: $$\lim _{a\to 0} 
F_{ext} =QB[-\delta(R)+\delta(R-S)]$$ 
This expression gives the following Newtonian eq. of motion for point-like 
particle
 $$m\dot v =QB[-\delta(R)+\delta(R-S)] =-QB {d \over d 
R}[-\theta(R)+\theta(R-S)]\eqno ( 4)$$ Equation ( 4) has the first integral:  
$$m{v^2 \over 2}+QB[\theta(R)-\theta(R-S)] = const =m{v_0^2 \over 2} $$ 
Thus 
for $0<R<S$ we have $$v=\sqrt{v_0^2 -(2QB/m)}$$ 
This solution describes 
the overcoming of a point-like particle the potential barrier only if the 
initial velocity $v_0$ is greater then the critical value $v_{cr}$, equal to 
$2QB/m$ or in dimensionless form $${v_{cr}\over c} \equiv \dot y_{cr} 
=\sqrt{2\lambda} \eqno( 5)$$

 If $k \equiv 0$ then the equation (3)  for $-1/2< y<+1/2,\ \ \ 
\ 0<\tau<t_1$ has the solution
$$y =-1/2 +v_0 \tau -\lambda \tau ^2/2,\ \ \ \ \dot y =v_0 -\lambda \tau 
\eqno(6)$$ It follows from (6)  that the point $y=+1/2$ is achieved (if the velocity 
 $\dot y $ is always positive)   at the moment of time $t_1$:  
$$t_1 ={v_0 -\sqrt{(v_0)^2 
-2\lambda} \over \lambda} \eqno (7)$$ 
Consequently the potential barrier is overcome if
 $v_0 >\sqrt{2\lambda}$, and the minimal initial velocity for this is
$v_0 =\sqrt{2\lambda}$ with $\dot y(t_1)=0$ and 
$$t_1 =v_0/\lambda =2/v_0 \eqno(8)$$
 (the positive part of the force in the equation (3) for $1/2< 
y<3/2$  does not hinder the motion of the particle, so
for analysis of conditions for overcoming the barrier one can consider 
only the interval:
 $-1/2< 
y<+1/2$ ),  

so for $t_1 =2$, one must take $v_0=1$ and
 $\lambda = 1/2$, and for  $t_1 =3$  --- 
$v_0= 2/3$ and $\lambda =2/9$ and so on.

If $k$ in (3) is not equal to zero and if the point $y=1/2$ 
is achieved at the moment of time $\tau=t_1$ in such a way that
 $\dot y (t_1)=0$ (this means that in the preceding moments of time the velocity 
is positive)
 then the further motion of particle,
 $1/2<y<3/2$,  is governed by the equation $${d^2 y \over d\tau 
^2} +k {d y(\tau ) \over d\tau }= k    {d y(\tau -1) \over d\tau }  + \lambda 
\equiv f$$ where $f>0$.

This eq. with the above condition $\dot y (t_1)=0$ has the solution
$$\dot y = e^{(-k\tau)}\int\limits_{t_1}^{\tau} dt' f(t')e^{(kt')}>0$$
consequently the barrier is overcome and the particle continues its motion 
for $\tau>t_1$.

Taking all this into consideration, lets construct the " tunneling" solutions
of the equation (3).

Let the Sommerfeld particle reach the point 
 $y=-1/2$ at the moment of time $\tau =0$, $y=1/2$ --- at $\tau =t_1$.

Let the particle's velocity for $\tau<0$ be constant:
$${d y \over d\tau }=v_0=const$$ 

Let the condition be true: $\dot y (t_1)=0$ (i.e. the barrier is overcome)
and let $$t_1 =2 \eqno (9)$$

Then one must integrate the eq. (3) into two stages: in the first, for the interval
$0<\tau<1,\ \ \ \ -1/2 <y<y_1$ and,in the  second, for the interval
 $1<\tau<2,\ \ \ \ y_1 <y<2$.
 
Solution on the first interval is
$$\dot y =v_0- {\lambda \over k}+{\lambda \over k} e^{(-k\tau)}\eqno (10)$$
and on the second --- 
$$\dot y =v_0- 2{\lambda \over k}+a_0 e^{(-kt)}+{\lambda \over k} \tau e^{(-k\tau)},$$
$$a_0= {\lambda \over k} (e^{(k\tau)}+1) -\lambda e^{(k\tau)} \eqno (11)$$
Conditions $y(t_1=2)=1/2$ ¨ $\dot y(t_1=2)=0$ are reduced to the set
$$v_0 =f_{1}\lambda$$
$$\lambda=1/f_{2}$$
here
$$f_1 ={2 \over k}- e^{(-k)}(1+{1 \over k}+{e^{(-k)} \over k}) \eqno 
(12)$$ $$f_2= {1-e^{(-k)} \over k} +{3-2e^{(-k)} -e^{(-2k)} \over 
k^2} -2e^{(-k)}(1+{1 \over k}+{e^{(-k)} \over k}) \eqno (13)$$
The analysis of the functions $f_1 (k),\ \ f_2 (k)$ (12, 13) shows that
the condition 
$v_0< \sqrt{2\lambda}$ for the solution (10, 11)
is valid due to the inequality 
$f_1 / \sqrt{2 f_2}<1$ - i.e. there is the tunneling - the initial velocity has
the forbidden, from newtonian point of view,  values, but ought to
the inequality $f_1/f_2 > 1$ all of them are greater then $1$:  $v_0>1$ ( for $k \equiv 0$, in accordance with 
(8), $v_0=1$ and $\lambda=1/2$ ).
 
Let now instead of (9) be
$$t_1=3 \eqno (14)$$

Then one must integrate the eq. (3) into tree stages: 
in the first, for the interval
 $0<\tau<1,\ \ \ \ -1/2 <y<y_1$ with the solution in the form (10), 
in the second, 
for the interval $1<\tau<2,\ \ \ \ y_1 <y<y_2$  with the solution in the form
 (11), and in the third, for the interval  $2<\tau<3,\ \ \ \ y_2 <y<1/2$ 
with the solution in the form
$$\dot y =v_0- 3{\lambda  \over k}+
e^{(-kt)}(a_1 +ke^{(k)}(a_0-\lambda e^{(k)} )\tau +k 
\lambda e^{(2k)} \tau ^2/2),$$
$$a_1 =e^{(2k)}({\lambda  \over k} -2\lambda+2k\lambda-\lambda e^{(-k)} 
+{\lambda  \over k}e^{(-k)}) \eqno (15)$$

Conditions $y(t_1=3)=1/2$ and $\dot y(t_1=3)=0$ are reduced to the
set
$$v_0 =f_{3}\lambda$$
$$\lambda=1/f_{4}$$
here
$$f_3 ={e^{(-3k)}\over 
2k}(6e^{(3k)}-(k^2+2k+2)e^{(2k)}-(4k+2)e^{(k)}-2) 
\eqno (16)$$ 
 
 $$f_4 ={e^{(-3k)}\over 
2k^2} 
 ((6k+12)e^{(3k)}-(3k^3+7k^2+10k+6)e^{(2k)}-(12k^2+10k+4)e^{(k)}-6k-2) 
 \eqno (17)$$ Analysis of functions (16, 17) shows that the condition
 $v_0< 
\sqrt{2\lambda}$ for the solution (15) is valid ---
 $f_3 / \sqrt{2 f_4}<1$ - i.e. there is the tunneling, and, 
contrary to the above case, the range of values of $v_0$ is: $2/3<v_0<1$
 (for $k \equiv 0$ 
 in accordance with (8), $v_0=2/3$ and $\lambda=2/9$ ).

The construction of tunneling solutions one can continue for
 $t_1 =4,\ 5,\ 
6,...$ and so on. It is obviously that the greater is the value of $t_1$,
the lower is the value of the minimal velocity $v_0$ necessary to 
overcome the barrier, but the more is complicated the form of the solution.

\eject
 \centerline {\bf{REFERENCES}}

  \begin{enumerate}
\item
 A.Sommerfeld, Gottingen Nachrichten, 29 (1904), 363 (1904), 201
  (1905).
\item
 L.Page, Phys.Rev., 11, 377 (1918)

 T.Erber, Fortschr. Phys., 9, 343 (1961)

 P.Pearle in "Electromagnetism",ed. D.Tepliz, (Plenum, N.Y.,
1982), p.211.

 A.Yaghjian, "Relativistic Dynamics of a Charged Sphere".
  Lecture Notes in Physics, 11 (Springer-Verlag, Berlin, 1992).

 F.Rohrlich, Am.J.Phys., 65(11), 1051 (1997). Phys.Rev., D60, 084017 (1999).
\item Alexander A. Vlasov, physics/9911059.
\end{enumerate}

\end{document}